\documentclass[a4paper,twoside]{article}

\usepackage{epsfig}
\usepackage{subcaption}
\usepackage{calc}
\usepackage{amssymb}
\usepackage{amstext}
\usepackage{amsmath}
\usepackage{amsthm}
\usepackage{multicol}
\usepackage{pslatex}
\usepackage{apalike}
\usepackage{algorithm2e}
\usepackage{url}
\usepackage[bottom]{footmisc}
\usepackage{SCITEPRESS}

\begin{document}

\title{Driving Through the Network: Performance and Workload Under Latency and Video Impairments}

 \author{\authorname{Ines Trautmannsheimer\sup{1}\orcidAuthor{0009-0004-0147-0183}, Ahmed Azab\sup{1} \orcidAuthor{0009-0008-1528-6511} and Frank Diermeyer\sup{1}\orcidAuthor{0000-0003-1441-5226}}
 \affiliation{\sup{1}Technical University of Munich, School of Engineering and Design, Institute of Automotive Technology and Munich Institute of Robotics and Machine Intelligence (MIRMI), 85748 Munich, Germany}
 \email{{ines.trautmannsheimer, ahmed.azab, diermeyer}@tum.de}}

\keywords{Teleoperation, Latency, Video quality, Physiology-aware adaptation, User study, Automotive applications}

\abstract{Teleoperation promises to extend the operational envelope of automated vehicles, yet it critically depends on network latency and video quality. We report a fixed-base driving-simulator study (N=25) with a $2\times2$ manipulation of added latency (100/300~ms) and bitrate (500/2000~kbit/s), plus a best-case baseline (0~ms added, 9000~kbit/s). We measured effective glass-to-glass (G2G) latency per condition (baseline $\approx$413~ms; effective totals $\approx$500–700~ms) and verified stable framerate and encoder settings. Multimodal measures covered performance (speed, steering reversals, crashes), oculomotor behavior (blink rate, fixation duration), physiology (RR interval, heart rate, skin conductance), and subjective workload. Latency and bitrate each increased operator load and modestly affected performance. Physiological measures (heart rate, RR interval) exhibited sub-additive interactions, whereas performance and oculomotor interactions were small or non-significant. Equivalence tests showed that 300~ms with 2000~kbit/s was velocity-equivalent to best-case (SESOI $\pm$2~km/h), while 300~ms with 500~kbit/s was not. We argue that latency and video quality should be treated as largely independent design levers, and that physiology-aware adaptation can anticipate overload before safety is compromised.}

\onecolumn \maketitle \normalsize \setcounter{footnote}{0} \vfill
\section{Introduction}
\begin{figure}[ht]
  \centering

  \subfloat[Start scenario]{\includegraphics[width=0.31\linewidth]{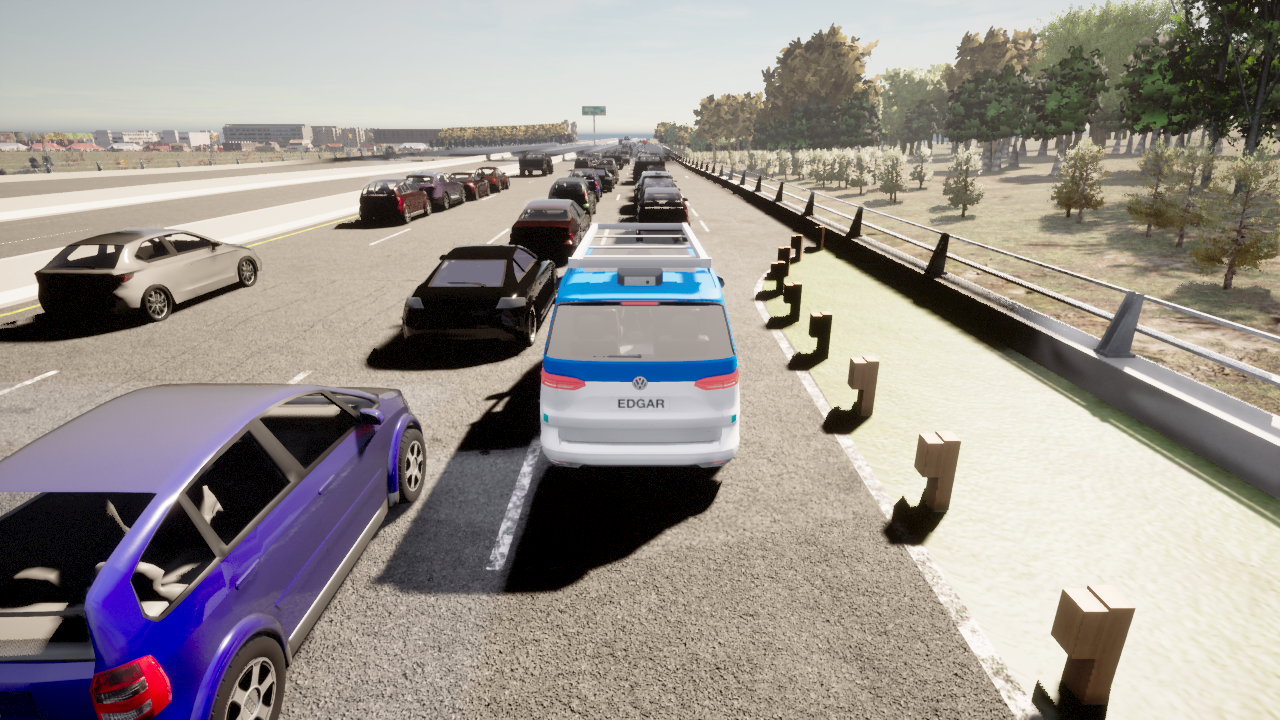}}\hfill
  \subfloat[Vehicle swerving]{\includegraphics[width=0.31\linewidth]{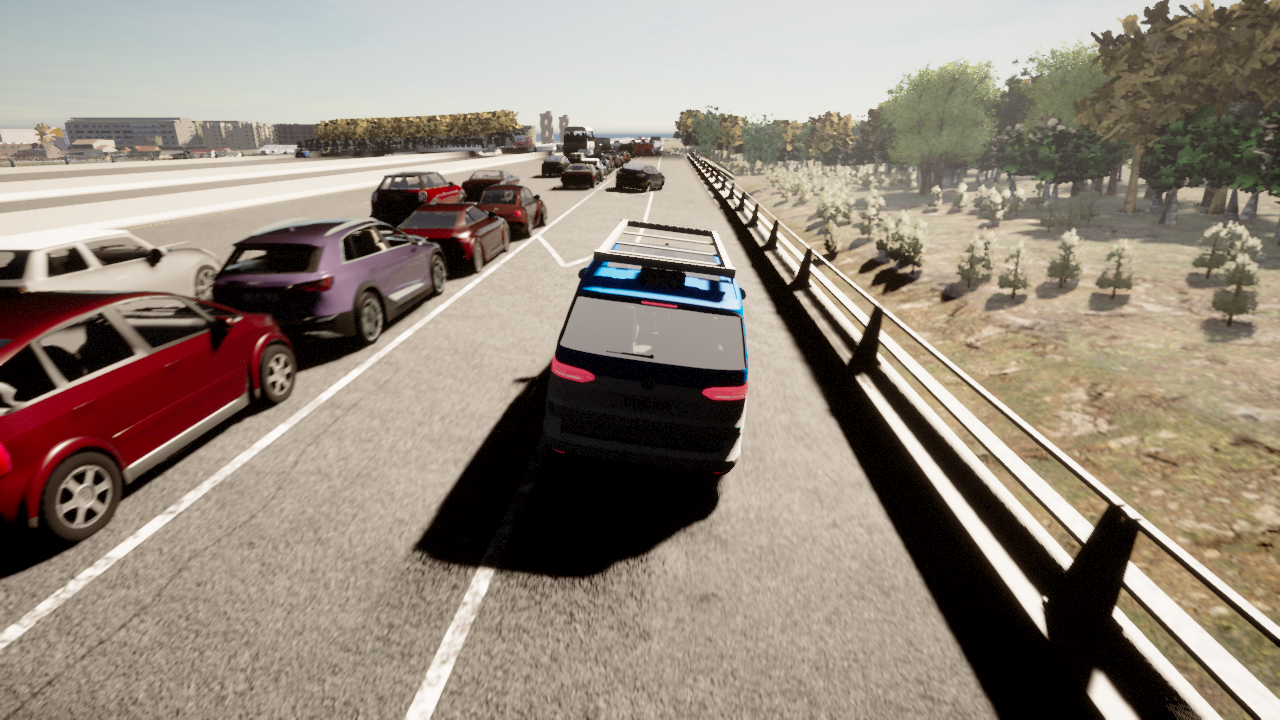}}\hfill
  \subfloat[Static obstacle (palm tree)]{\includegraphics[width=0.31\linewidth]{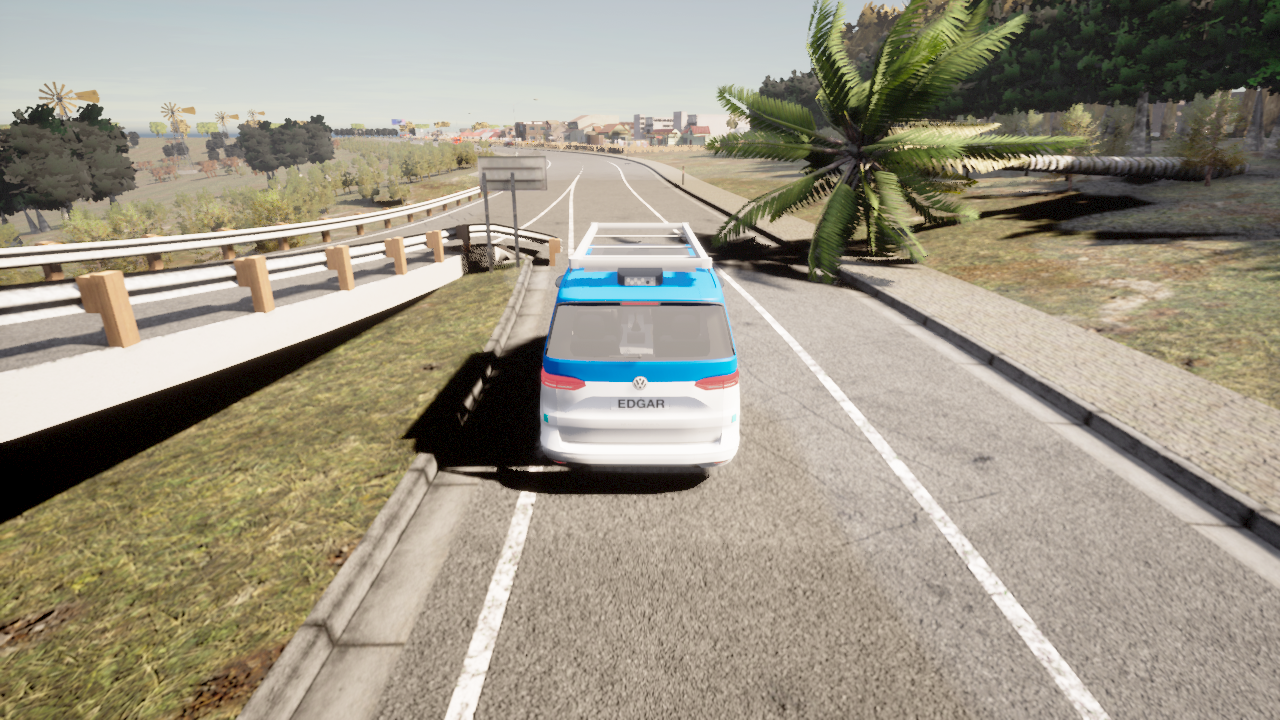}}\\

  \subfloat[Speed bump]{\includegraphics[width=0.31\linewidth]{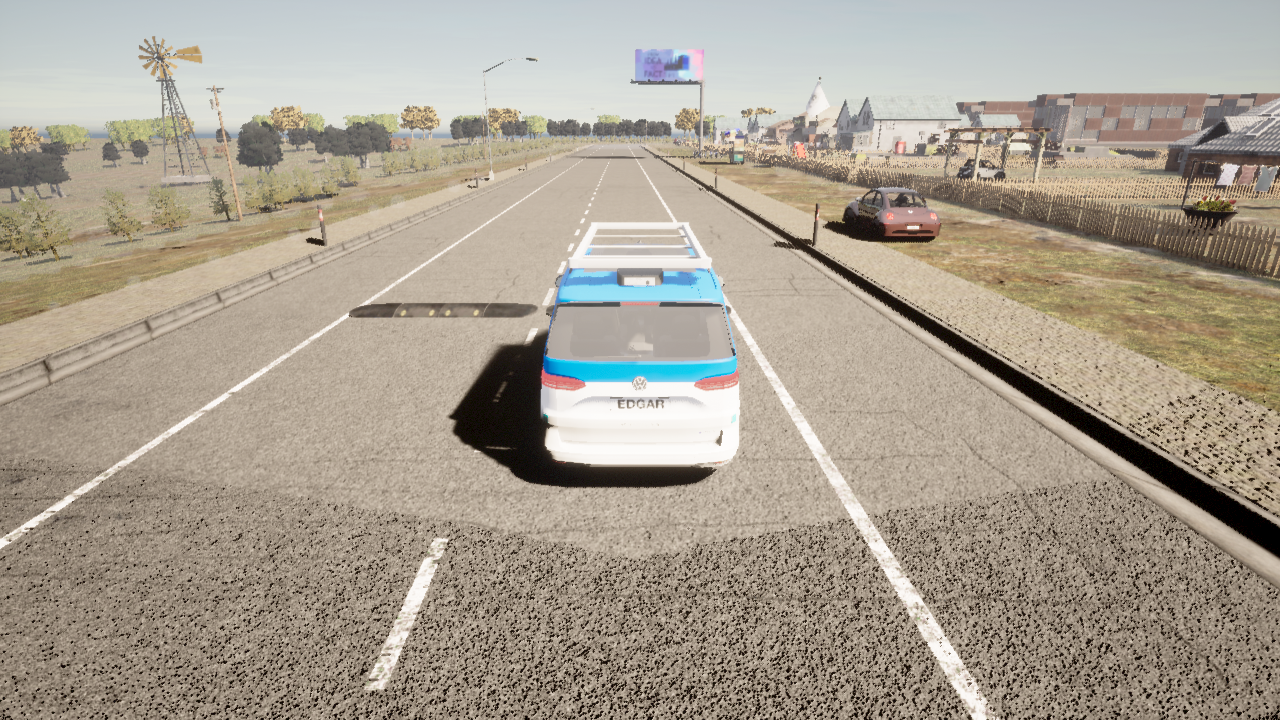}}\hfill
  \subfloat[Rolling box]{\includegraphics[width=0.31\linewidth]{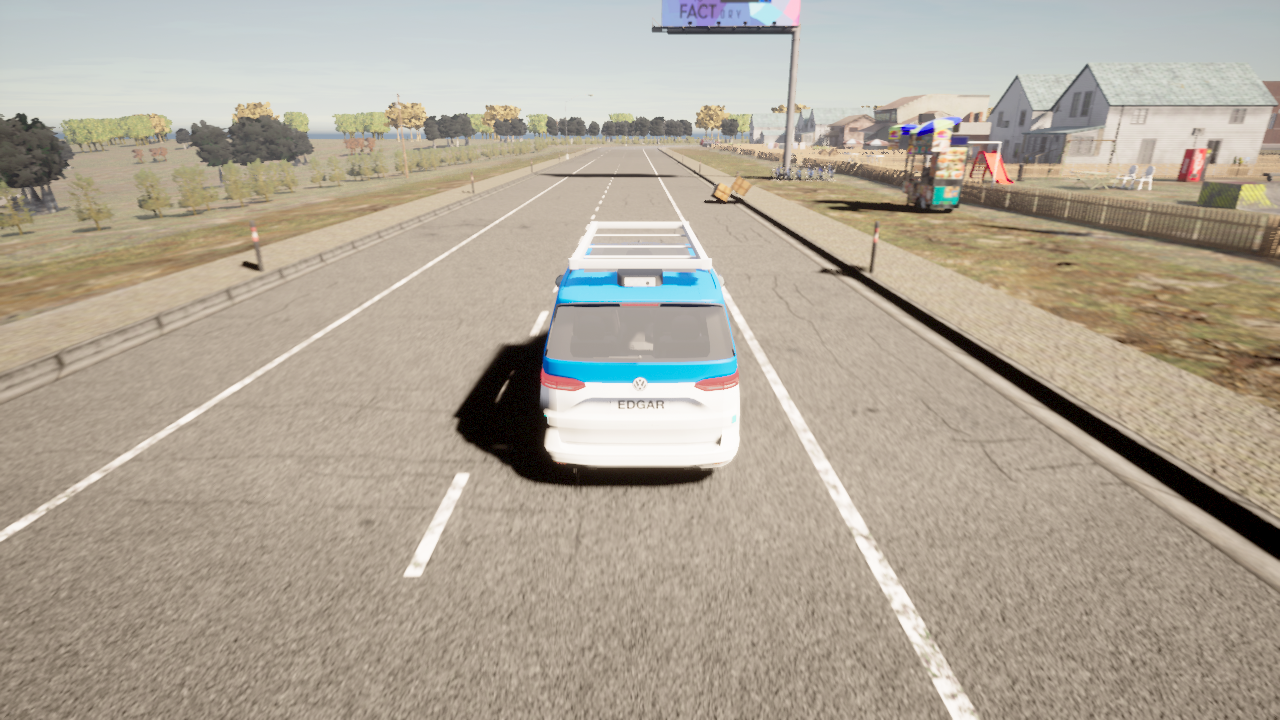}}\hfill
  \subfloat[Static obstacle (trash)]{\includegraphics[width=0.31\linewidth]{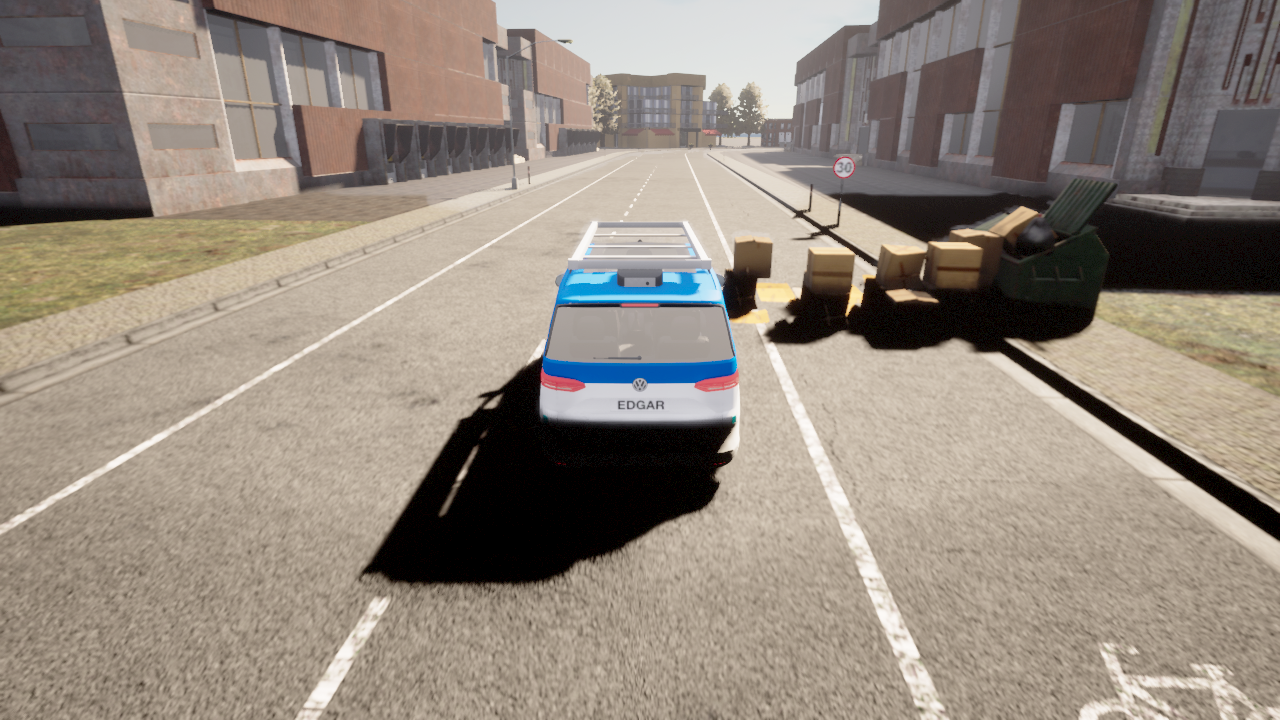}}\\

  \subfloat[Construction site]{\includegraphics[width=0.31\linewidth]{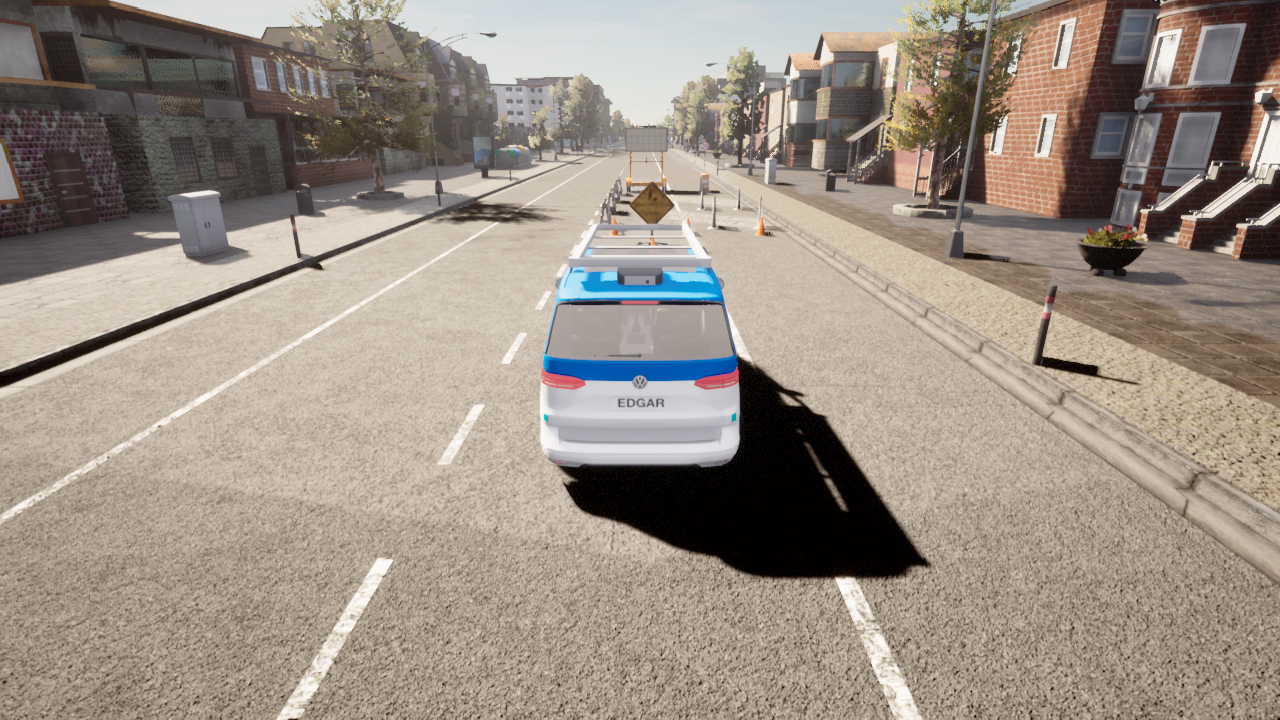}}\hfill
  \subfloat[Crosswalk]{\includegraphics[width=0.31\linewidth]{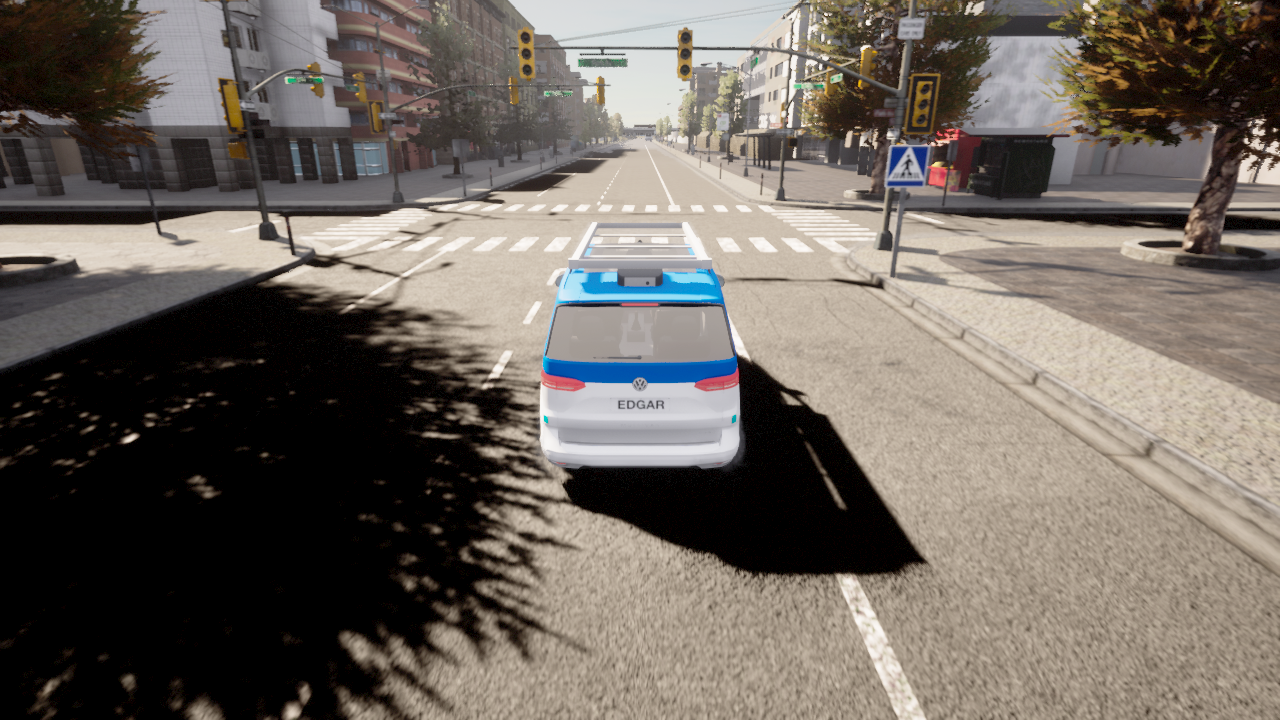}}\hfill
  \subfloat[Pedestrians crossing]{\includegraphics[width=0.31\linewidth]{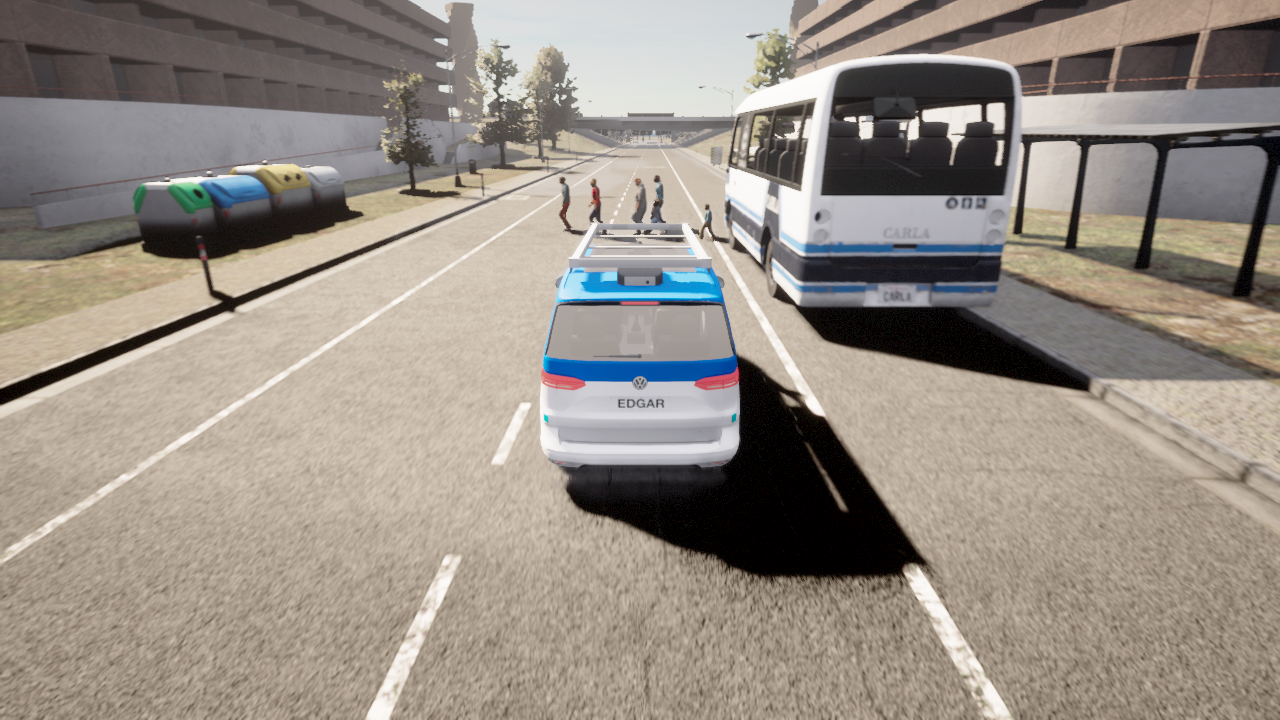}}

  \caption{Overview of representative teleoperation scenarios used in the study.}
  \label{fig:example scenarios}
\end{figure}

Autonomous driving has transitioned from a long-standing vision of science fiction to a rapidly maturing technological reality. Companies such as Waymo \cite{Waymo2025DC} and Zoox \cite{Zoox2025Vegas} have already demonstrated that fully driverless operation is technically feasible. Yet even with this progress, automated vehicles still encounter situations that exceed their Operational Design Domain and require human judgment or situational awareness that current AI systems cannot yet replicate.

Teleoperation offers a scalable solution to these limitations: instead of having a human driver on site, a remote operator supervises and, when necessary, directly controls vehicles from a distance \cite{brecht2024evaluation}. This concept allows one operator to assist multiple vehicles sequentially, improving flexibility and reducing costs. However, it also introduces a critical dependency on networked visual communication. The operator’s situational awareness is entirely mediated by video streams transmitted over mobile networks, where limited bandwidth and variable latency can severely affect perception, control, and decision-making.

Previous research has mainly examined these two factors—latency and video quality—in isolation, focusing either on the perceptual consequences of compression or on control delays introduced by network lag. In real teleoperation, however, both impairments often occur simultaneously. Their combined impact on operator workload and performance, and the implications for intelligent interface design, remain poorly understood.

This study addresses this gap by systematically investigating the joint effects of latency and video degradation in a controlled simulator experiment. Participants remotely operated a vehicle under varying network conditions while physiological, behavioral, and subjective measures were recorded. Our goal is to identify how these factors interact and how insights from multimodal measurements can inform the design of physiology-aware, adaptive user interfaces that anticipate operator overload before safety is compromised.

\subsection{Connected Vehicles and Teleoperation}
\label{sec:teleoperation}


A vehicle is designated "connected" if it constantly communicates with its environment. A variety of connections exist between vehicles and the environment. 
This mobile connection enables the remote operator to establish a connection to the vehicle \cite{lichiardopol2007survey} and thus receive and send data. For example, this can be camera data, point clouds, or object lists \cite{kerbl2025tum}. The operator then uses this information to build up situational awareness and sends control signals back to the vehicle, which then implements them and sends feedback back to the operator in the form of new data. 
The control signals can be of different types; there are different control concepts that enable the operator to control the vehicle. These are divided into two categories: Remote Driving and Remote Assistance \cite{majstorovic2022survey}. In this study, we will focus on direct control, which is part of remote driving. This means that the operator takes control using input devices that are modeled on a real vehicle. They can set the steering angle with the steering wheel and the speed with the brakes and pedals. These are then transmitted directly to the vehicle and then adapted.

\subsection{Latency in Teleoperation of Connected Vehicles}
\label{sec:latency}
As Zhao et al. \cite{zhao_remote_2024} have previously determined in their review of existing literature on teleoperation, latency constitutes one of the three primary challenges, alongside remote driving feedback and support control. They also mention the effects of latency on performance, emotional states and mental workload. In order to comprehend these influences, it is necessary to examine two key stages in the teleoperation control loop as outlined in section \ref{sec:teleoperation}. The transmission of data from the vehicle to the operator and subsequently back to the vehicle is mandated. Data transmission can be categorized into two primary classifications: cellular and non-cellular. The term "cellular" is a comprehensive designation encompassing all mobile technologies and standards, including 3G, 4G (LTE), and 5G. The non-cellular category encompasses a range of wireless technologies, including LAN, WLAN, WiFi, Bluetooth, and satellite.
Each transmission type exhibits distinct latencies and data rates, in addition to specific ranges. In the non-cellular sector, LAN/WiFi or WLAN has low latencies of 50~ms, a data rate of 500 Mbps, and a coverage range of 10 to 300 meters. However, they are not suitable for use in connected vehicles due to their limited range \cite{kamtam2024network}. Measurements were also carried out in the USA. Xiao et al. \cite{xiao_driving_2019} conducted urban driving tests with LTE and obtained an average value of 81.9~ms, with a standard deviation of 29.5~ms. However, a separate study conducted by the OECD in 2025 \cite{oecd_state_connectivity_2025} in the United States revealed latency values of up to 358~ms.

Presently, LTE remains a pertinent technology for the transmission of data from connected vehicles, as the availability of 5G coverage is inadequate, particularly within Germany. As indicated by Kerbl et al. \cite{kerbl2025tum}, LTE is employed in the software configuration. The findings demonstrated that LTE resulted in a median of 36~ms more compared to LAN, culminating in a glass-to-glass latency of 150 to 200~ms when incorporating the latency of the video processing pipeline and rendering.
The consequences of this additional latency were investigated by Neumeier et al. \cite{neumeier2019teleoperation} investigated the consequences of this additional latency in various situations in their study. They chose three different levels: 0~ms, 150~ms, and 300~ms. They combined these with different scenarios, such as a pylon scenario, parking, or a double curve. The analysis shows that a latency of 150~ms did not significantly impair driving performance compared to no latency, whereas 300~ms led to clear performance degradation, particularly in complex scenarios like the Pylon task. In contrast, in simpler or low-speed scenarios, even 300~ms latency was often tolerable. 
In their study on the teleoperation of unmanned ground vehicles, Luck et al. \cite{luck2006An} investigated different directions of latency and duration at different levels of vehicle autonomy. The findings indicated that the duration of latency exerted a substantial influence on completion time, yet no significant impact was observed on driving performance. However, the direction of latency did not influence the results and was not subjectively perceived as a difference. It was determined that the complexity of the task being performed by the robot is a pivotal factor in the overall assessment. In instances where the task is deemed to be elementary, the latency issue becomes less significant. Conversely, in more challenging scenarios, such as navigation in confined spaces, latency emerges as a pivotal factor. 

\subsection{Effects of Image Quality}

As already described in section \ref{sec:latency}, there is often only a limited bandwidth available in teleoperation for LTE, which is on average 3 - 8 ~Mbit/s \cite{kamtam2024network}. This means that the video must be heavily compressed in order to be transmitted to the operator. Depending on the codec, this results in losses in quality and artifacts the less bandwidth is available.
Both lossless and lossy reductions are possible. The video may be precisely recreated at the receiver to match the original thanks to lossless compression, which only eliminates statistical redundancies. Nevertheless, this technique fails to attain a compression ratio high enough for effective transmission at a lower data rate. Lossy compression is therefore employed in contemporary technology. Significantly larger compression ratios and, thus, more data reduction are made possible by this technique \cite{richardson2004h}.
Hoffmann et al. \cite{hoffmann2022quantifying} conducted a study on the effects of this reduced video quality in teleoperation, in which they artificially degraded the image quality with the help of various quantizers and measured the reaction times in different situations.
The results reveal significant effects of image quality on both subjective and objective measures. While quality levels Q44 and Q50 led to significantly longer reaction times and were judged as unsuitable for teleoperation due to safety concerns, Q36 marked the threshold at which participants first perceived a subjective degradation in IQ and task performance—despite no significant objective performance drop—suggesting that operators may adopt a conservative stance based on perceived quality, especially in critical scenarios.

\section{Study}
The objective of this study is to examine the interaction between latency and image quality, as these phenomena frequently co-occur with compromised mobile communications quality. In order to achieve the above-mentioned objective, a series of hypotheses were formulated. These were subsequently tested in the study. The data recorded during the process is then utilized in the evaluation to test the hypotheses. 
\subsection{Design \& Hypothesis}

Our design aims to measure both driving performance and subjective workload, thereby quantifying the influence of image quality and latency in teleoperation. To this end, we employed objective and physiological measures, specifically eye tracking data, skin conductance, the number of collisions, and task completion time. In addition, we collected subjective data using the NASA-TLX \cite{hartDevelopmentNASATLXTask1988} and supplementary questionnaire items. The following research questions guided the study:

\begin{itemize}
    \item \textbf{RQ1:} Does the simultaneous degradation of latency and image quality lead to a more than additive increase in workload compared to the isolated effects?

    \item \textbf{RQ2:} Does the simultaneous degradation of latency and image quality lead to a more than additive degradation of performance compared to the isolated effects?

    \item \textbf{RQ3:} Can differences in the sensitivity of individual measurement methods be identified with regard to combination effects?
    
\end{itemize}
The experimental design was a $2 \times 2$ factorial design, complemented by a best-case baseline condition. The nominal additional latency levels were set to 0, 100, and 300~ms, while image quality was manipulated via bitrate (9000, 2000, and 500~kbit/s), employing the H.264 codec \cite{ITU-H264} to induce compression-related degradations. 

Importantly, the 100 and 300~ms conditions refer to \emph{additionally induced} network delays. A round-trip manipulation check using a steering-angle injection revealed that the baseline system already incurred a mean glass-to-glass latency of 412.93~ms. Accordingly, the effective total latencies were approximately 500–700~ms across the experimental conditions.
The factorial design was embedded into a custom-built driving environment, ensuring high control and reproducibility.

\subsection{Construction of the scenarios}
\begin{figure}[t]
\centering
\subfloat[]{\includegraphics[width=0.32\linewidth]{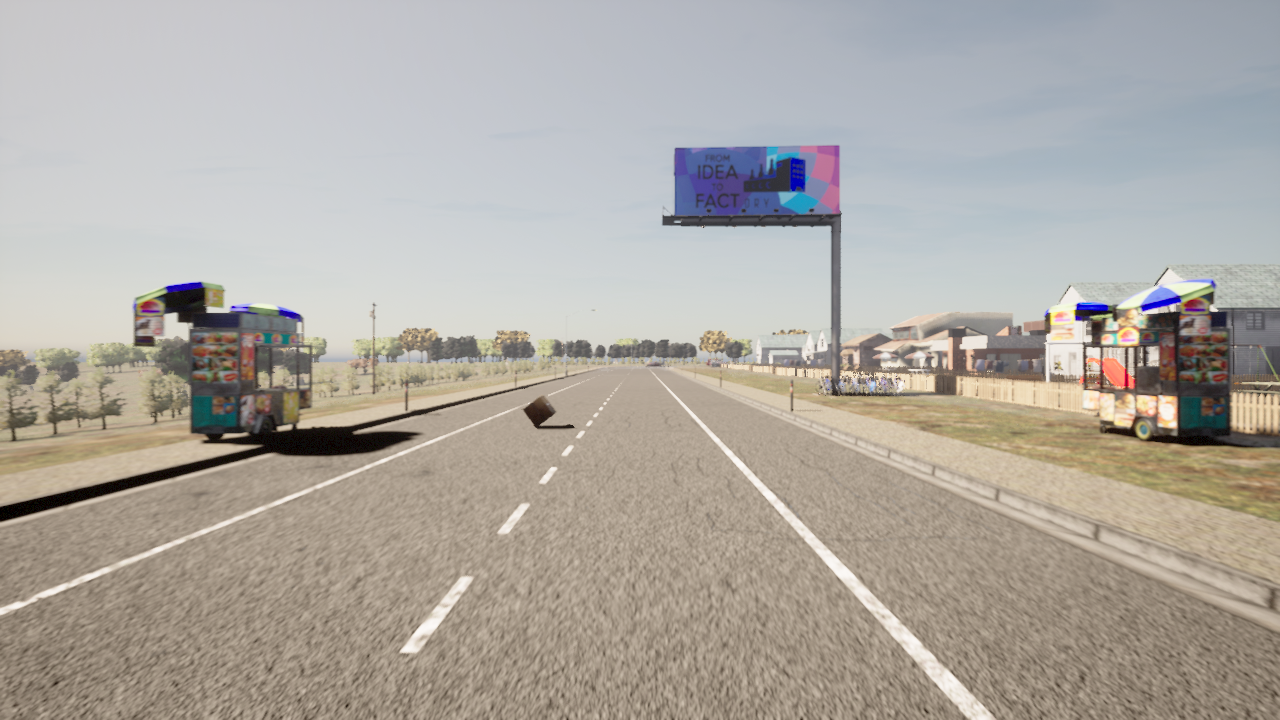}\label{fig:55}}\hfill
\subfloat[]{\includegraphics[width=0.32\linewidth]{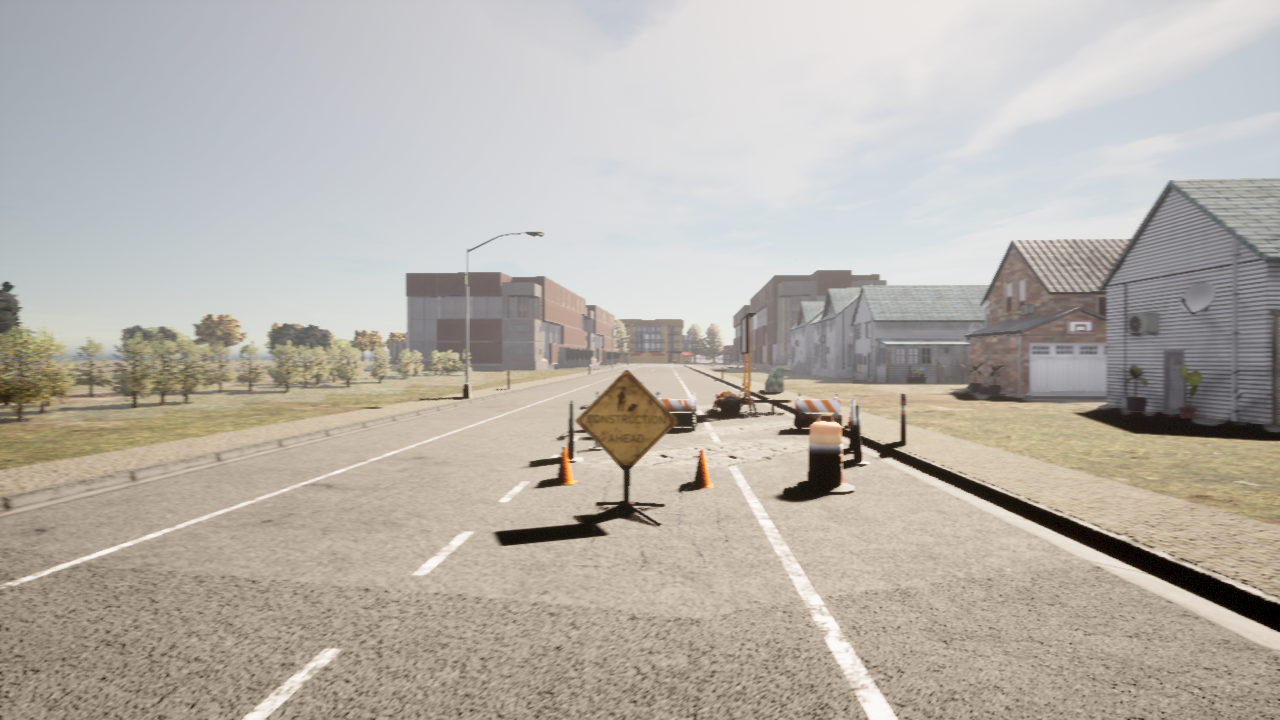}\label{fig:46}}\hfill
\subfloat[]{\includegraphics[width=0.32\linewidth]{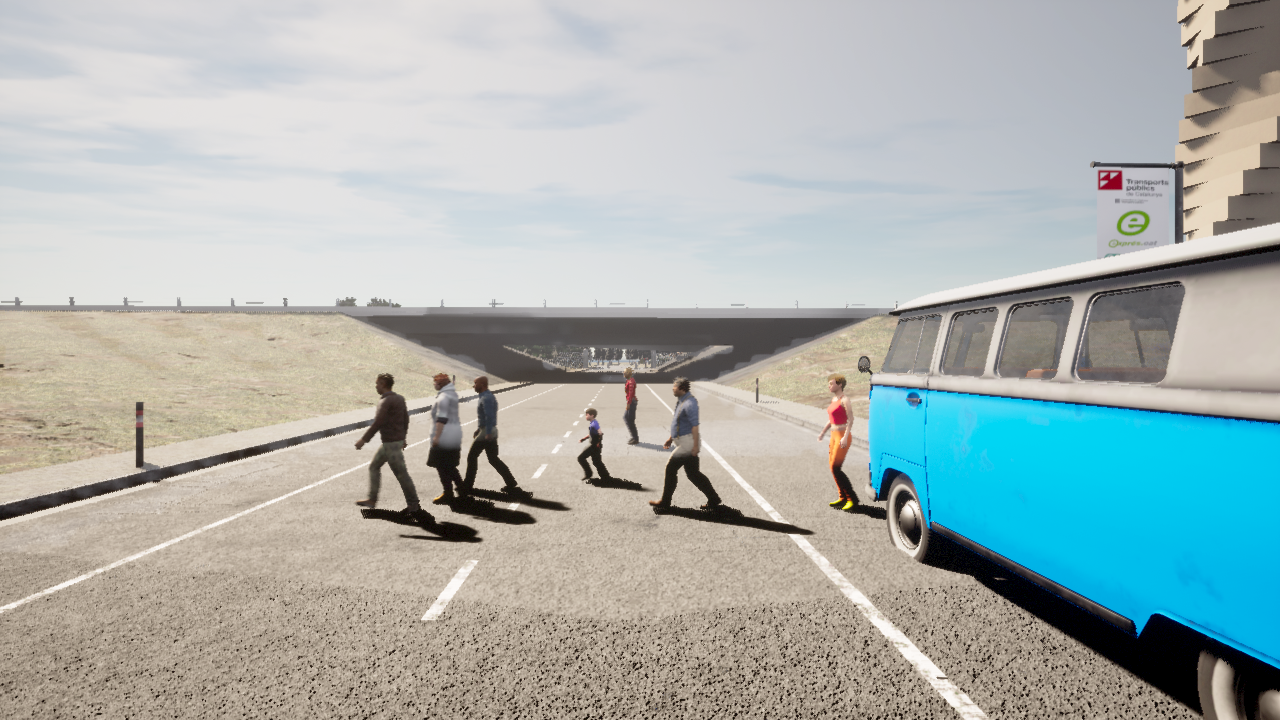}\label{fig:39}}

    \caption{Pictures of the nine scenarios, as perceived by the participants from the vehicle position.  Figure \ref{fig:55} displays the rolling box. Figure \ref{fig:46} presents a variant of the construction site, while Figure \ref{fig:39} shows a version of the pedestrians disembarking from the bus.}
    \label{fig:Scenarios vehicle persepective}
\end{figure}
A central methodological contribution of this study is the development of a bespoke test track specifically designed to reproduce disengagement scenarios. To replicate real-world conditions and create authentic challenges, we designed a custom road network and scenario set grounded in prior literature. Disengagements occur when an ODD boundary is reached or when the Automated Driving System malfunctions \cite{brecht2024evaluation}. According to the SAE J3016 \cite{on2021taxonomy}, vehicles operating at SAE Level 4 or 5 must transition to a minimal risk condition before deactivating automated control. In such a state, the vehicle comes to a safe stop but cannot resume its mission without external intervention. A remote operator can then partially or fully assume the dynamic driving task to restore functionality or guide the vehicle back into its ODD. The scenarios implemented in our custom track, covering failures in object detection, path planning, and trajectory execution, were selected to represent the most common classes of disengagements identified in the literature \cite{brecht2024evaluation}.

We choose a realistic scenario for evasion, in which an error in route planning is represented. On the motorway, the CAV cannot find a passable route to drive past a stationary car. The reason for this is a traffic jam caused by an accident that happened at the next exit, causing the CAV behind the stationary car to come to a complete stop. To bypass the roadblock, the vehicle must leave its ODD, which in this case are motorways. Therefore, the RO must take action, leave the motorway at the next exit and drive around the detour until it returns to the next motorway entrance in the same direction and rejoins the ODD. 

\begin{figure}[!htbp]%

	\centering%
	\includegraphics[width=\linewidth]{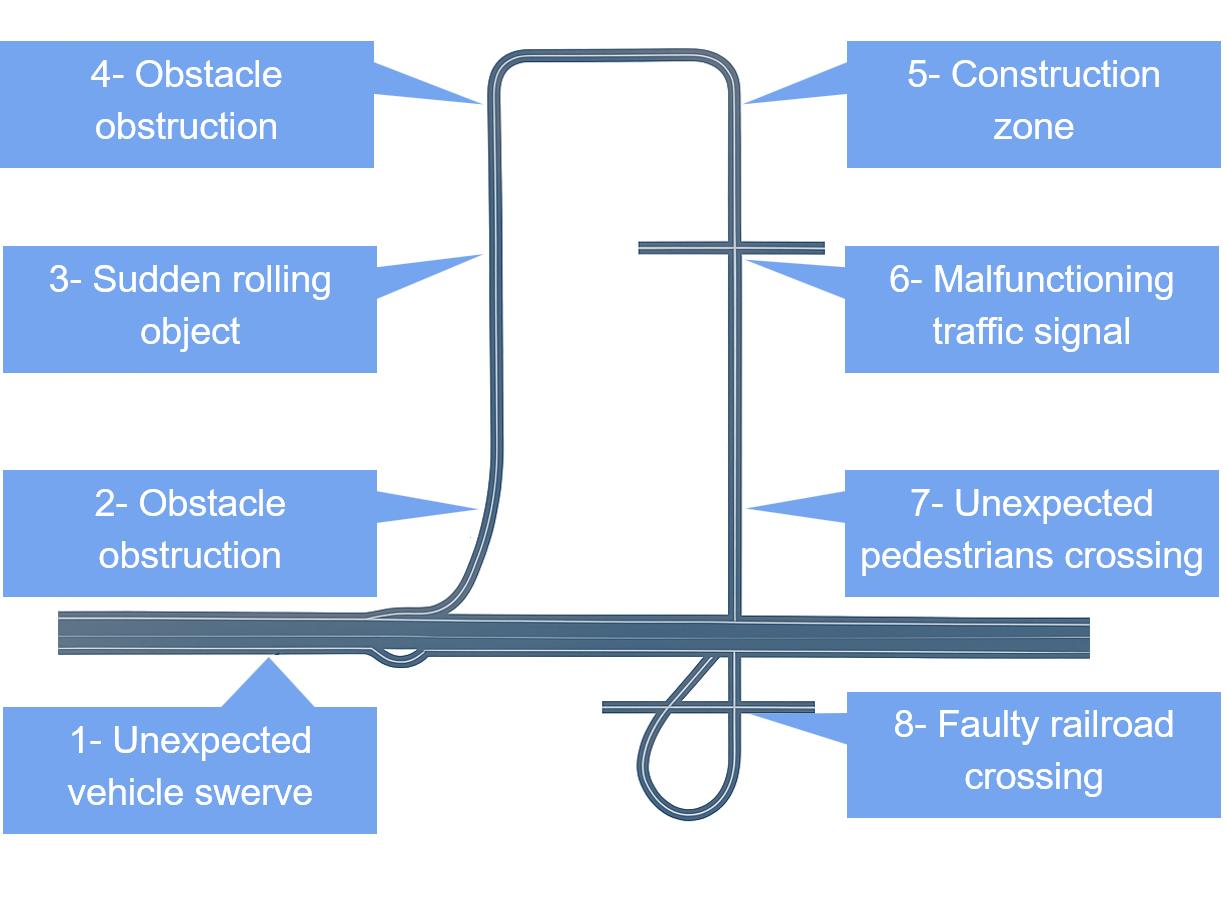}%
	\caption{Overview of the route geometry and shows, using one of our maps as an example, how the events were distributed.}
	\label{fig:Events location overview}%
\end{figure}

If teleoperation support were always initiated at the same point, unforeseen events could prevent remote operators from re-engaging automation, ultimately resulting in CAV failure. To avoid such dead ends and maintain experimental control, we implemented all disengagement scenarios within a single custom-built map. Events were placed in short succession, allowing operators to complete one scenario, briefly return to normal mode, and immediately proceed to the next. This design eliminated the need to restart separate maps for each scenario, which would have required over 30 distinct runs and frequent container restarts, causing long waiting times and reduced participant engagement. By consolidating all scenarios into one cohesive track, we minimized overhead, preserved ecological validity, and ensured that participants remained focused throughout demanding takeover events. This design choice was critical for reliably assessing how latency and image quality affect both driving performance and operator workload.
Examples of scenarios from the participant's perspective can be seen in Figure~\ref{fig:Scenarios vehicle persepective}.
The following scenarios were included in our track and distributed along the route, as can be seen in Figure~\ref{fig:Events location overview}:
\begin{itemize}
\item Unexpected vehicle swerve 
\item Obstacle obstruction (Fallen Tree, Full Trash Bin, ...)
\item Road construction zone \cite{brecht2024evaluation}, \cite{richter2023components}
\item Sudden Rolling Box \cite{hoffmann2022quantifying}
\item Malfunctioning traffic signal
\item Unexpected pedestrian crossing \cite{richter2023components}, \cite{saparia2021active}
\item Faulty railroad crossing (lights + gates)
\end{itemize}
Using MathWorks RoadRunner \cite{matlab_roadrunner}, we designed the underlying road network and traffic infrastructure. The track begins on a highway, includes an exit at segment 2, and re-enters the highway at segment 9, with the experiment concluding at segment 10. Between these points, the disengagement events were positioned. Two intersections (segments 6 and 8) were created to host the malfunctioning traffic signal and the faulty railroad crossing scenarios.

The completed road network was exported and imported into our simulation environment, CARLA \cite{Dosovitskiy17}, \cite{malik2022carla}, and further developed in Unreal Engine \cite{unrealengine}. Unreal provided a broad asset library to build a realistic environment; additional elements, such as the railroad crossing package, were procured externally and successfully integrated. To ensure immersion, the placement of all events was carefully designed to balance ecological validity and experimental control (see Figure~\ref{fig:Events location overview}).

Dynamic events including unexpected pedestrian crossings, rolling obstacles, and sudden vehicle swerves were scripted using the CARLA Python API \cite{carla_python_api}. For each of the five maps, dedicated scripts were created to trigger events consistently during study execution. In the case of the malfunctioning railroad crossing, the gate was animated to cycle between 20 seconds open and 10 seconds closed, providing a reproducible but realistic failure case.

\subsection{Setup}
The video encoding and streaming were implemented using GStreamer (version~1.22; \cite{GStreamer}) inside a Docker-based environment. 
We employed the standard plugin set (base/good/bad/ugly, libav) including \texttt{x264enc} for H.264 encoding in compliance with ITU-T H.264/AVC \cite{ITU-H264}. 
This setup ensured reproducibility across different machines and allowed precise control of both latency and bitrate. 
We quantified glass-to-glass (G2G) latency using a steering-angle injection and timestamp alignment, reporting mean±SD per cell. To validate the bitrate manipulation, framerate (20 Hz) and encoder configuration  were held constant.

The software setup was initialized using a pre-scripted software routine, and all sensor systems, including the GSR tablet, the eye-tracking system, and the electrocardiogram (ECG, Polar H10), were prepared and tested in advance.
The study was conducted using a Docker-based infrastructure. The simulation environment and the control software were divided and managed flexibly using numerous Docker containers.
The complete environment, inclusive of the generated maps, was incorporated within the initial Docker container designed for the CARLA simulation program.
The vehicle simulation 
software, which serves as the link between the control software and the simulation software, was housed in the second Docker container.
The  program that interacted with the simulation was housed in the third Docker container, which contained the operator and the vehicle. This container facilitated the configuration and control of the connections to the operator and the simulated vehicle. Furthermore, it demonstrated the capacity to dynamically modify critical parameters while the simulation was in progress. Through the utilization of the interfaces, the frame rate settings employed through the video manager panel and the latency settings managed by the latency manager panel were directly associated with the program, thereby facilitating flexible control throughout the experimental trials.
All measurement devices were synchronized with the simulation framework inside Docker, ensuring reproducibility.

\subsection{Apparatus}
\begin{figure}[!htbp]%
	\centering%
	\includegraphics[width=\linewidth]{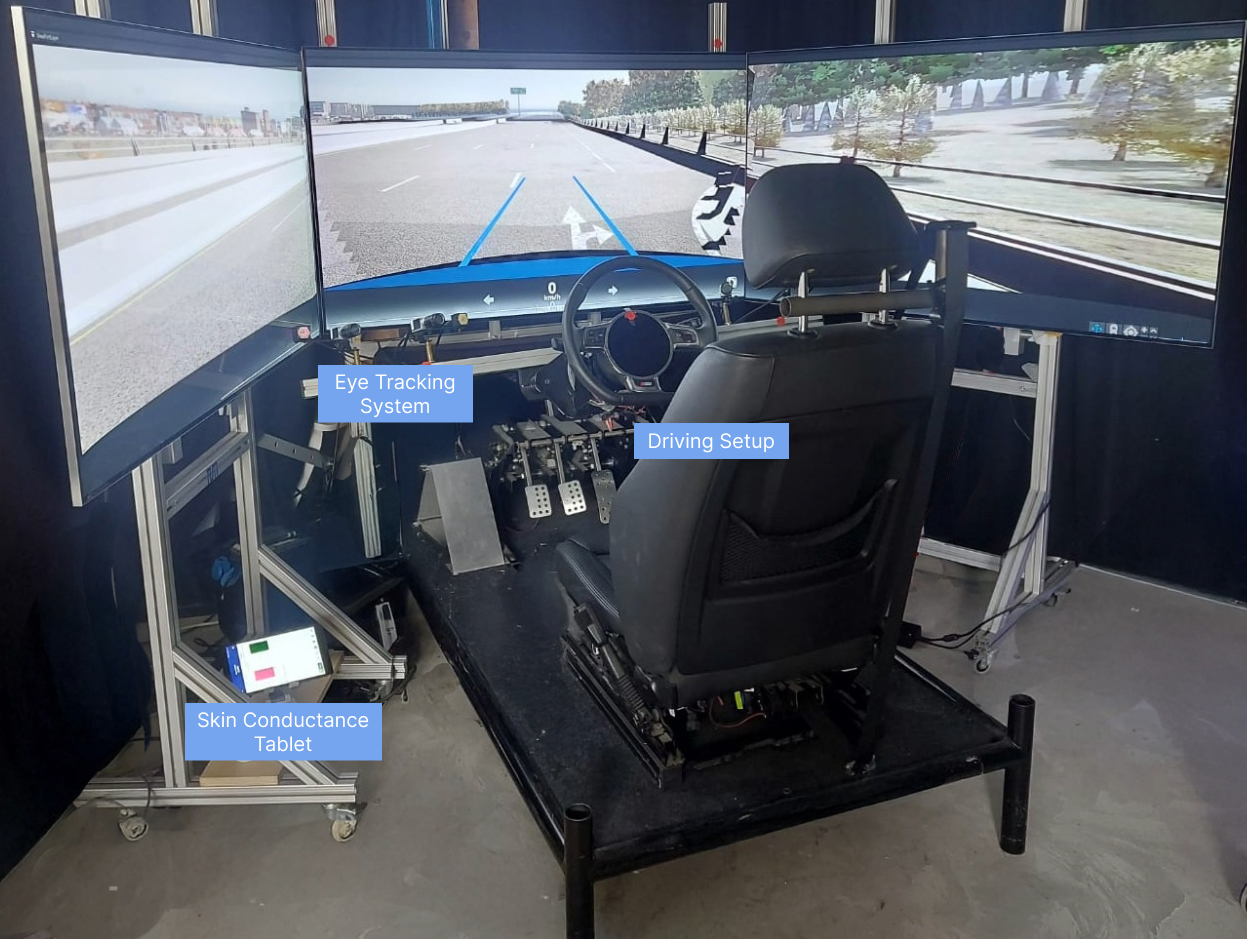}%
	\caption{The experimental setup incorporated a workstation positioned beneath the screens of the Eye Tracking System, along with the tablet utilized for skin conductance measurement, which can be connected to the electrodes. The driving setup comprised an authentic car seat, a steering wheel equipped with paddle shifters to facilitate gear changes, and pedals (throttle and brake).}
	\label{fig:Setup}%
\end{figure}%

As illustrated in Figure \ref{fig:Setup}, the workstation features three 55-inch LG OLED screens with 4K resolution, in addition to a specially constructed static seat box equipped with a steering wheel and pedals. The front left camera is displayed on the left screen, the right camera is displayed on the right, and the middle screen displays the camera at the car. The seat's adjustability is comparable to that of a conventional automobile seat.

The system, developed by Smart Eye, incorporated three infrared flashes and four cameras positioned horizontally behind the steering wheel. This configuration enables the recording of eye motions from all required perspectives, ensuring precise tracking. A negligible variation of merely 0.11–0.16 pixels per camera was obtained through camera calibration.

The skin conductance device is characterized by its simplicity in terms of setup, which involves the use of a Huawei tablet, the eSense application \cite{mindfield_skinresponse}, and measuring sensors from eSense, which include two electrodes and can be linked to skin pads. Given that both hands would be utilized for steering and the right leg would be used for the throttle and brake, the attachment would be placed on the left foot before logging for each participant. The temporal data, in the form of the clock time, and the microSiemens value are meticulously documented in the logging process. Subsequent to the conclusion of the logging process, a CSV file is created, encompassing the entirety of the participant's logged data.

The Polar H10 \cite{polar_h10} has been utilized as an ECG  measured device that is equipped with the Polar Pro chest strap. The device's compatibility with PCs that are equipped with a Bluetooth module hinges upon the utilization of Bluetooth as the primary means of data transfer. A Python script has been developed to facilitate the creation of a graphical user interface for the purpose of controlling the device. The device is initially connected via Bluetooth to the Windows system, thereby assigning each participant a unique identifier and trial number. The RR-interval, UNIX time, and the device's logging are all included in the aforementioned study.

\subsection{Procedure}
Participants received a standardized briefing and provided written informed consent before instrumentation. A short familiarization drive introduced the simulator interface and the measurement devices (eye tracking, skin conductance, ECG); no experimental data were recorded during familiarization.

Each participant then completed five drives. Each drive followed a fixed route on a custom map that contained  multiple scripted disengagement scenarios (e.g., rolling obstacle, pedestrian crossing, construction zone). Scenario order and map variants were counterbalanced across drives to reduce anticipation and learning. After each drive, participants completed a computerized questionnaire (NASA-TLX subscales and additional items). Trial order (drive index) and scenario identifiers were logged to support statistical control of order and route-segment effects.

The experiment was conducted in a dark, quiet room. Sensors were calibrated prior to the first drive (eye-tracking calibration) and checked for signal quality between drives. At the end, participants completed a short debriefing questionnaire and were de-instrumented.
\begin{figure}[t]
  \centering
  \includegraphics[width=\linewidth]{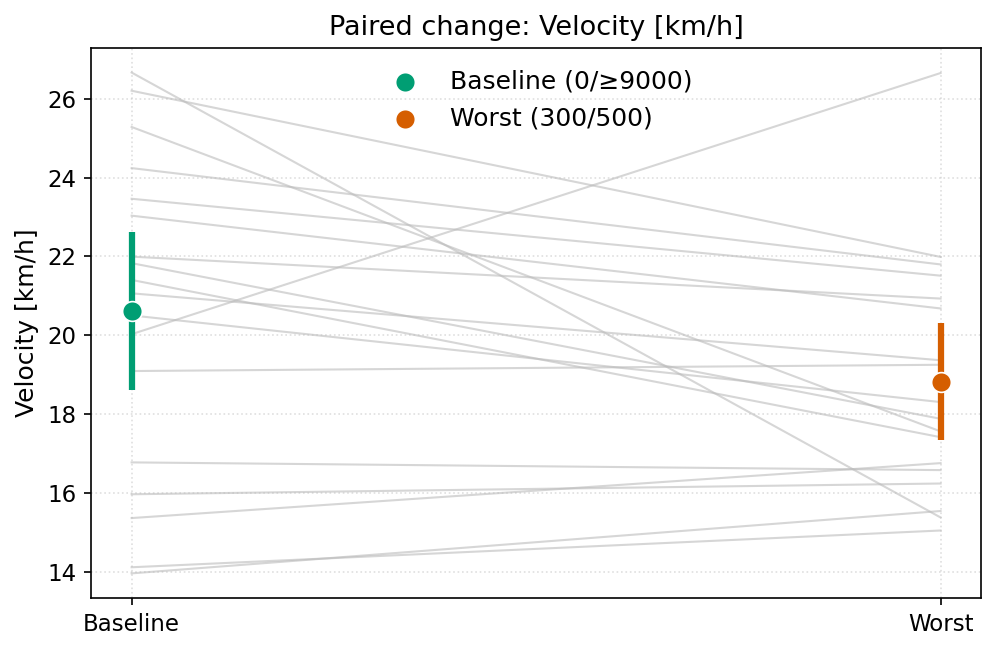}
  \caption{Paired comparison for Velocity between Best-Case (0~ms, $\geq$9000~kbit/s) and Worst-Case (300~ms, 500~kbit/s). 
  Thin gray lines show individual participants; markers and bars show the mean $\pm$ 95\% CI. 
  Velocity is significantly lower in the Worst-Case (Wilcoxon $p < .05$).}
  \label{fig:paired-vel}
\end{figure}

\subsection{Measurements}
We collected a set of multi-modal variables: crash frequency, SRR, task completion time, eye tracking features (blink rate, pupil diameter, fixation duration), skin conductance, RR-interval, perceived workload (raw NASA TLX \cite{hartDevelopmentNASATLXTask1988}). SRR refers to the number of times a driver changes the direction of the steering wheel per minute.

\subsection{Manipulation Checks}
\label{sec:manipchecks}
To validate the experimental factors, we quantified effective glass-to-glass (G2G) latency and objective video quality.

\paragraph{G2G latency.} We used a steering-angle injection with timestamp alignment to estimate end-to-end latency from operator input to on-screen visual feedback. For each cell of the $2\times2$ design. The best-case pipeline (0~ms added, 9000~kbit/s) yielded a baseline of approximately 413~ms; adding 100/300~ms delays produced effective totals of roughly 500–700~ms. Within-cell variability was low.

\paragraph{Video quality and stability.} The video pipeline ran on a single workstation to minimize network variance. Apart from the manipulated factors (added delay, bitrate), framerate (20~Hz) and encoder configuration (H.264/x264, constant settings for preset, GOP/key-interval, and rate control) were held constant across conditions. To corroborate bitrate as a proxy for image quality, we computed PSNR/SSIM/VMAF on short representative clips from each condition; quality scaled monotonically with bitrate. A framerate stability check confirmed negligible drift across conditions.

\paragraph{Implication.} These checks indicate that observed effects can be attributed to the intended manipulations rather than uncontrolled jitter or encoder fluctuations.
\begin{figure}[t]
  \centering
  \includegraphics[width=\linewidth]{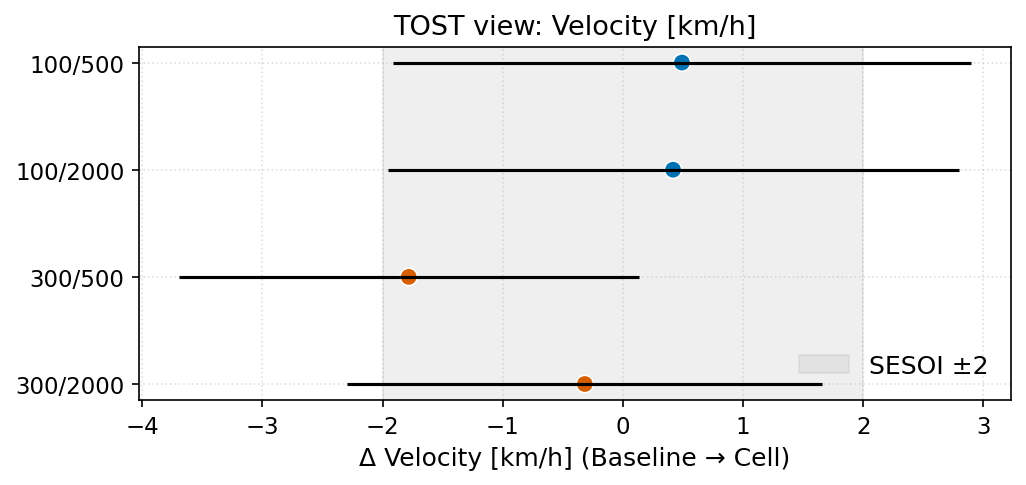}
  \caption{Difference vs. Best-Case for Velocity (Cell $-$ Best-Case). 
  Points show mean paired differences; horizontal lines show 95\% CIs. 
  The shaded band depicts the equivalence region (SESOI $\pm$2~km/h). 
  The 300/2000 cell falls within the equivalence region, indicating equivalence to Best-Case, whereas 300/500 does not.}
  \label{fig:tost-vel}
\end{figure}
\subsection{Participants}
Initially, 28 participants took part in the study; however, one participant was deemed ineligible due to his inability to operate a vehicle without wearing corrective glasses, a requirement for the study's eye-tracking system. The data from two participants were excluded from the analysis. The first participant completed ten test drives, rather than the designated five. As a result, the study duration, which was approximately two and a half to three hours, was deemed excessively prolonged. Consequently, commencing with the second participant, the number of test drives was reduced to five. It was observed that another participant demonstrated a lack of seriousness in their driving, and they did not familiarize themselves with the simulator interface. Consequently, the final analysis encompasses data from 25 participants. The mean age of the participants was 27.5 years (SD=3.5), with a minimum of 20 and a maximum of 38 years.
The comparatively young age distribution is attributed to the recruitment of participants primarily from a university environment. Nevertheless, this age range is considered realistic, as future remote operators are expected to be young and technologically literate \cite{neumeier2019teleoperation}. Among the 25 participants, 21 identified as male and 4 as female.

\section{Results}
Table~\ref{tab:mixedlm_results} summarizes the outcomes across performance, oculomotor, and physiological measures. Consistent with the interaction plots, performance and oculomotor measures yielded only small effects without reliable interactions; by contrast, heart rate and RR interval showed robust main effects and significant sub-additive interactions (with skin conductance trending similarly).




\begin{figure}[t]
  \centering
  \includegraphics[width=\linewidth]{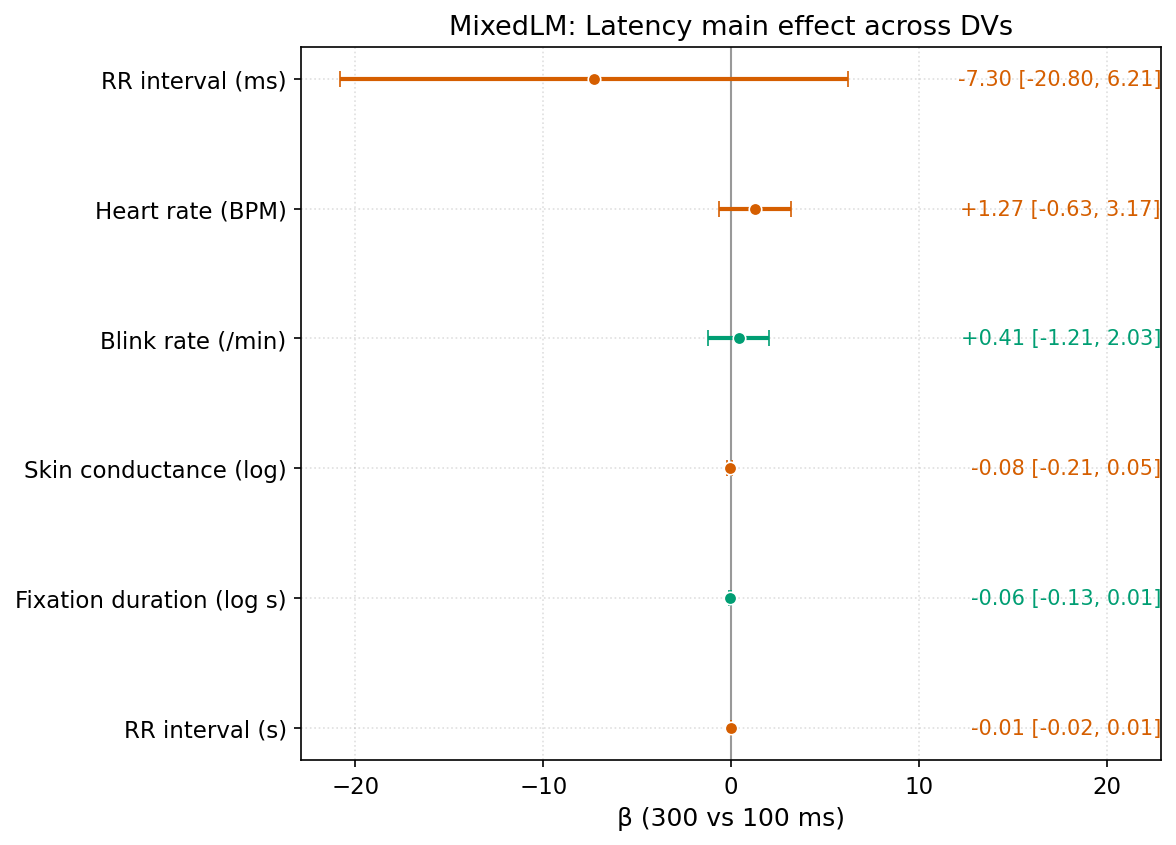}
  \caption{Mixed-effects fixed-effect estimates ($\beta$) and 95\% CIs for the latency main effect (300 vs. 100~ms) across all dependent variables. 
  Colors encode domains (Performance, Oculomotor, Physiology). 
  Physiological measures (heart rate, RR interval) show robust sensitivity, whereas performance and oculomotor measures exhibit smaller, non-significant changes.}
  \label{fig:forest-latency}
\end{figure}

\subsection{NASA-TLX}
Subjective workload ratings showed stable distributions across latency and image quality (all one-way ANOVAs $p > .05$). 
When comparing Baseline (0~ms, 9000~kbit/s) to the most degraded condition (300~ms, 500~kbit/s), Mental Demand and Effort increased significantly (both $p < .05$). 
No other subscales showed systematic differences. Overall, operators reported consistently high Mental Demand and Frustration, while Performance was perceived as least concerning.

\begin{table*}[t]
\centering
\caption{Model results for all dependent variables except crashes. Crash counts were analysed separately with clustered GEE (Table~\ref{tab:crashes_gee}). 
Reference: Latency = 100\,ms, Image Quality = 500\,kbit/s. 
Significance: * $p<.05$, ** $p<.01$, *** $p<.001$, $\dagger$ $p<.10$.}
\label{tab:mixedlm_results}

\setlength{\tabcolsep}{3pt} 
\renewcommand{\arraystretch}{1.1}

\begin{tabular}{lcccc}
\hline
 & \multicolumn{3}{c}{\textbf{Effects}} & \\
\cline{2-4}
\textbf{Dependent variable} 
& \textbf{Latency} 
& \textbf{Quality} 
& \textbf{Lat.×Qual.} 
& \textbf{Intercept} \\
\hline
Velocity [km/h]                 
& $-3.20$ (n.s.)   
& $-1.98$ (n.s.)   
& $+3.27$ (n.s.)    
& $22.83^{***}$ \\

Skin conductance [log $\mu$S]   
& $-0.01$ (n.s.)   
& $-0.02$ (n.s.)   
& $+0.21^{\dagger}$ 
& $1.73^{***}$ \\

Heart rate [BPM]                
& $+12.2^{***}$    
& $+14.0^{***}$    
& $-13.2^{***}$     
& $65.73^{***}$ \\

RR interval [s]                 
& $+0.13^{***}$    
& $+0.12^{***}$    
& $-0.13^{***}$     
& $0.65^{***}$ \\

Blink rate [/min]               
& $+0.59$ (n.s.)   
& $-0.01$ (n.s.)   
& $-0.15$ (n.s.)    
& $8.19^{***}$ \\

Fixation duration [log s]       
& $-0.06^{\dagger}$
& $-0.04$ (n.s.)   
& $+0.01$ (n.s.)    
& $-0.91^{***}$ \\

Steering reversal rate [log1p/min] 
& $+0.09^{*}$   
& $+0.04$ (n.s.)   
& $+0.21^{\dagger}$ 
& $0.03$ (n.s.) \\

\hline
\end{tabular}
\end{table*}
\subsection{Performance Measures}
Mean velocity decreased under higher latency and lower image quality, but without statistical significance. 
Steering reversal rate (SRR) increased with 300~ms latency ($p < .05$), and crashes trended higher in degraded conditions. 
The 300/2000 condition was statistically equivalent to Baseline in velocity (TOST within $\pm 2$~km/h), whereas 300/500 was not.
Crash counts were analysed with Generalized Estimating Equations (GEE; participant-clustered robust SE). 
A Poisson model indicated a higher, yet non-significant, crash rate at 300~ms latency relative to 100~ms 
(IRR $=2.29$, 95\%~CI $[0.69,\,7.55]$, $p=.175$). Image quality showed no reliable effect 
(2000 vs.\ 500~kbit/s: IRR $=0.86$, 95\%~CI $[0.22,\,3.30]$, $p=.823$), and the interaction was negligible 
(IRR $=1.09$, 95\%~CI $[0.21,\,5.73]$, $p=.916$). 
Results were virtually identical when using a Negative Binomial family. 
The baseline crash rate in the reference cell (100~ms, 500~kbit/s) was $\exp(-1.273)\approx 0.28$ crashes per trial.

\begin{table}[h!]
\centering
\caption{Crash counts (per trial): GEE Poisson with participant-clustered robust SE. Entries are incidence rate ratios (IRR) with 95\% CIs. Results were virtually identical with a Negative Binomial family. Reference cell: Latency = 100~ms, Image Quality = 500~kbit/s.}
\begin{tabular}{lrr}
\hline
Predictor & IRR [95\% CI] & $p$ \\
\hline
Latency (300 vs.~100~ms) & 2.29 [0.69, 7.55] & .175 \\
Quality (2000 vs.~500~kbit/s) & 0.86 [0.22, 3.30] & .823 \\
Latency $\times$ Quality & 1.09 [0.21, 5.73] & .916 \\
\hline
\end{tabular}
\label{tab:crashes_gee}
\end{table}

\subsection{Oculomotor Measures}
Blink rate and fixation duration showed only weak, non-significant trends. 
Oculomotor responses were not robust to the manipulations compared to other modalities.

\subsection{Physiological Measures}
Physiological signals were most sensitive. 
Heart rate increased under both higher latency and lower image quality, with significant interaction terms indicating non-additive effects. 
RR intervals shortened accordingly, consistent with elevated arousal. 
Skin conductance showed a trend-level interaction, suggesting cumulative sympathetic activation when latency and quality impairments co-occurred.

\section{Discussion}
In a controlled simulator with measured G2G latency and stable encoding, both added delay and reduced bitrate independently increased operator load and modestly degraded performance. Importantly, physiological measures exhibited \emph{sub-additive} interactions, while performance and oculomotor interactions were small or non-significant. 
These results provide insight into how human operators adapt to degraded visual communication and where intelligent support interfaces could intervene before performance deteriorates. 

\subsection{Performance under high latency budgets}
Average speed declined with higher delay and lower bitrate, but effects were small once the already elevated baseline G2G latency was considered. This aligns with threshold accounts: beyond a certain budget, further increments yield diminishing marginal harm on speed. Nevertheless, crashes trended upward under 300~ms added delay, underscoring that tail risks can grow even when mean performance shifts are modest. Steering reversal rate (SRR) showed a small but reliable latency effect, indicating that fine-grained control corrections become more frequent under delay.
From an interface-design perspective, such compensation behavior suggests that feedback channels—visual or haptic—could be dynamically tuned when latency exceeds individual tolerance thresholds, thereby stabilizing control effort.

\subsection{Workload: self-report vs.\ physiology}
NASA-TLX captured differences primarily between best-case and the most degraded cell; intermediate manipulations produced relatively stable ratings. Physiology (heart rate, RR interval) responded consistently to both factors and showed \emph{sub-additive} interactions: combined impairments did not exceed the sum of single-factor effects, suggesting constrained headroom rather than runaway overload. Skin conductance provided trend-level support. Together, these findings suggest that self-report underestimates early overload, while physiology offers a more sensitive channel for online monitoring.

This pattern highlights the potential of physiological signals as implicit input for intelligent teleoperation UIs.  
By continuously tracking HR and RR, a system could infer increasing load and adapt visual density, cue timing, or information rate accordingly—representing a key step toward closed-loop operator support.

\subsection{Design levers and acceptable operating regions}
Equivalence tests complemented null-hypothesis testing: 300~ms with 2000~kbit/s remained velocity-equivalent to best-case within a $\pm$2~km/h SESOI, whereas 300~ms with 500~kbit/s did not. Practically, latency and bitrate can be treated as largely independent levers, but designers should avoid combinations that breach physiological headroom. 

These levers define the operating envelope within which adaptive policies can safely vary video parameters or interface complexity.  
Our results motivate physiology-aware adaptation (e.g., escalating assistance or visual simplification when RR/HR drift into risk bands) before overt performance loss.

\subsection{Methodological guidance}
First, manipulation checks matter: reporting measured, not only configured, latency/quality reduces interpretability gaps. Second, route/segment structure influences speed; accounting for Scenario/Segment (fixed or random effects) or within-scenario standardization avoids confounding. Third, multi-domain outcomes call for clear primary vs. secondary endpoints and FDR control.

Beyond methodological rigor, transparent measurement pipelines also enable future intelligent systems to align perceptual metrics with human-state indicators, improving data-driven adaptation models.

\subsection{Limitations and external validity}
Internal validity was high (single-host pipeline; measured G2G; stable encoder), but external validity is limited relative to cellular/WAN conditions with handovers and uplink variability. The baseline G2G latency was substantial, potentially compressing headroom for detecting interactions in performance outcomes. Crashes were rare, limiting power for safety endpoints. 
Future work should integrate real network variability and test how adaptive UIs can mitigate these fluctuations in situ.  
This includes evaluating live 4G/5G handovers, diversifying operator populations, and implementing real-time physiology-driven feedback loops to evaluate usability, trust, and transparency.

\section{Conclusion}
Added delay and reduced bitrate independently increased operator load and modestly affected performance in a fixed-base driving simulator with measured G2G latency and stable encoding. Physiological measures (heart rate, RR interval) revealed sub-additive interactions, whereas performance and oculomotor interactions were small or absent. Equivalence testing identified operating points that remained acceptable (e.g., 300~ms with 2000~kbit/s) and others that did not (300~ms with 500~kbit/s). 

These findings delineate boundaries for adaptive teleoperation interfaces.  
They show that latency and video quality can be treated as separate but monitorable design dimensions, each informing the level of system assistance.  
We recommend treating latency and video quality as largely independent design levers, instrumenting teleoperation with physiology-aware monitoring to anticipate overload, and reporting manipulation checks to strengthen interpretability across studies.  
Ultimately, integrating these insights into intelligent user interfaces may enable remote driving systems that adapt to the operator’s cognitive and physiological state in real time, enhancing both safety and trust.

\section*{Ethical Considerations}

This study was conducted in full compliance with institutional and GDPR data-protection regulations. 
The experimental protocol was reviewed and approved by the responsible non-medical ethics committee of the technical university of Munich. 

\section*{Acknowledgments}

Ines Trautmannsheimer conceived the study idea and developed the research concept. 
She supervised the project and contributed to the experimental design and analysis. 
Ahmed Azab implemented the technical system, conducted the study, and contributed to the data analysis and manuscript preparation. Frank Diermeyer made essential contributions to the conception of the research projects and
revised the paper critically for important intellectual
content. He gave final approval for the version to be
published and agrees to all aspects of the work. As
a guarantor, he accepts responsibility for the overall
integrity of the paper. The research was financially
supported by the Federal Ministry of Research, Tech-
nology and Space of Germany (BMFTR) within the
project ASUR, No. 03ZU2105BA.

\bibliographystyle{apalike}
{\small
\bibliography{references}

\begin{thebibliography}{}

\bibitem[Brecht et~al., 2024]{brecht2024evaluation}
Brecht, D., Gehrke, N., Kerbl, T., Krauss, N., Majstorovi{\'c}, D., Pfab, F.,
  Wolf, M.-M., and Diermeyer, F. (2024).
\newblock Evaluation of teleoperation concepts to solve automated vehicle
  disengagements.
\newblock {\em IEEE Open Journal of Intelligent Transportation Systems}.

\bibitem[{CARLA Simulator Team}, 2025]{carla_python_api}
{CARLA Simulator Team} (2025).
\newblock {\em CARLA Simulator: Python API Reference}.
\newblock CARLA.
\newblock Accessed: 14 July 2025.

\bibitem[Committee, 2021]{on2021taxonomy}
Committee, O.-R. A. D.~O. (2021).
\newblock {\em Taxonomy and definitions for terms related to driving automation
  systems for on-road motor vehicles}.
\newblock SAE international.

\bibitem[Dosovitskiy et~al., 2017]{Dosovitskiy17}
Dosovitskiy, A., Ros, G., Codevilla, F., Lopez, A., and Koltun, V. (2017).
\newblock {CARLA}: {An} open urban driving simulator.
\newblock In {\em Proceedings of the 1st Annual Conference on Robot Learning},
  pages 1--16.

\bibitem[{Epic Games}, ]{unrealengine}
{Epic Games}.
\newblock Unreal engine.

\bibitem[{GStreamer Project}, ]{GStreamer}
{GStreamer Project}.
\newblock Gstreamer.
\newblock Accessed September 2025.

\bibitem[Hart and Staveland, 1988]{hartDevelopmentNASATLXTask1988}
Hart, S.~G. and Staveland, L.~E. (1988).
\newblock Development of {{NASA-TLX}} ({{Task Load Index}}): {{Results}} of
  {{Empirical}} and {{Theoretical Research}}.
\newblock In Hancock, P.~A. and Meshkati, N., editors, {\em Advances in
  {{Psychology}}}, volume~52 of {\em Human {{Mental Workload}}}, pages
  139--183. North-Holland.

\bibitem[Hoffmann et~al., 2022]{hoffmann2022quantifying}
Hoffmann, S., Willert, F., Hofbauer, M., Schimpe, A., and Diermeyer, F. (2022).
\newblock Quantifying the influence of image quality on operator reaction times
  for teleoperated road vehicles.
\newblock In {\em 13th International Conference on Applied Human Factors and
  Ergonomics (AHFE 2022)}.

\bibitem[{ITU-T}, 2019]{ITU-H264}
{ITU-T} (2019).
\newblock Advanced video coding for generic audiovisual services (h.264).
\newblock Technical Report Recommendation H.264, International
  Telecommunication Union.
\newblock Version approved February 2019.

\bibitem[Kamtam et~al., 2024]{kamtam2024network}
Kamtam, S.~B., Lu, Q., Bouali, F., Haas, O.~C., and Birrell, S. (2024).
\newblock Network latency in teleoperation of connected and autonomous
  vehicles: A review of trends, challenges, and mitigation strategies.
\newblock {\em Sensors (Basel, Switzerland)}, 24(12):3957.

\bibitem[Kerbl et~al., 2025]{kerbl2025tum}
Kerbl, T., Brecht, D., Gehrke, N., Karunainayagam, N., Krauss, N., Pfab, F.,
  Taupitz, R., Trautmannsheimer, I., Su, X., Wolf, M.-M., et~al. (2025).
\newblock Tum teleoperation: Open source software for remote driving and
  assistance of automated vehicles.
\newblock {\em arXiv preprint arXiv:2506.13933}.

\bibitem[Lichiardopol, 2007]{lichiardopol2007survey}
Lichiardopol, S. (2007).
\newblock A survey on teleoperation.

\bibitem[Luck et~al., 2006]{luck2006An}
Luck, J.~P., McDermott, P.~L., Allender, L., and Russell, D.~C. (2006).
\newblock An investigation of real world control of robotic assets under
  communication latency.
\newblock In {\em Proceedings of the 1st ACM SIGCHI/SIGART Conference on
  Human-Robot Interaction}, HRI '06, page 202–209, New York, NY, USA.
  Association for Computing Machinery.

\bibitem[Majstorovi{\'c} et~al., 2022]{majstorovic2022survey}
Majstorovi{\'c}, D., Hoffmann, S., Pfab, F., Schimpe, A., Wolf, M.-M., and
  Diermeyer, F. (2022).
\newblock Survey on teleoperation concepts for automated vehicles.
\newblock In {\em 2022 IEEE international conference on systems, man, and
  cybernetics (SMC)}, pages 1290--1296. IEEE.

\bibitem[Malik et~al., 2022]{malik2022carla}
Malik, S., Khan, M.~A., and El-Sayed, H. (2022).
\newblock Carla: Car learning to act—an inside out.
\newblock {\em Procedia Computer Science}, 198:742--749.

\bibitem[{MathWorks}, 2025]{matlab_roadrunner}
{MathWorks} (2025).
\newblock Roadrunner – 3d-editor zum erstellen von szenen für das
  simulations- und testen von automatisierten fahrsystemen.
\newblock Accessed on 11 September 2025.

\bibitem[{Mindfield Biosystems Ltd.}, 2025]{mindfield_skinresponse}
{Mindfield Biosystems Ltd.} (2025).
\newblock Skin response [apparatus and software].
\newblock Accessed: 21 August 2025.

\bibitem[Neumeier et~al., 2019]{neumeier2019teleoperation}
Neumeier, S., Wintersberger, P., Frison, A.-K., Becher, A., Facchi, C., and
  Riener, A. (2019).
\newblock Teleoperation: The holy grail to solve problems of automated driving?
  sure, but latency matters.
\newblock In {\em Proceedings of the 11th International Conference on
  Automotive User Interfaces and Interactive Vehicular Applications},
  AutomotiveUI '19, page 186–197, New York, NY, USA. Association for
  Computing Machinery.

\bibitem[{OECD}, 2025]{oecd_state_connectivity_2025}
{OECD} (2025).
\newblock State of connectivity between and within countries: Closing broadband
  connectivity divides for all.
\newblock Technical report, Organisation for Economic Co-operation and
  Development (OECD).
\newblock Accessed on 11 Sep 2025.

\bibitem[{Polar Electro Oy}, 2025]{polar_h10}
{Polar Electro Oy} (2025).
\newblock Polar h10 heart rate sensor.
\newblock Accessed: 21 August 2025.

\bibitem[Richardson, 2004]{richardson2004h}
Richardson, I.~E. (2004).
\newblock {\em H. 264 and MPEG-4 video compression: video coding for
  next-generation multimedia}.
\newblock John Wiley \& Sons.

\bibitem[Richter et~al., 2023]{richter2023components}
Richter, A., Walz, T.~P., Dhanani, M., H{\"a}ring, I., Vogelbacher, G.,
  H{\"o}flinger, F., Finger, J., and Stolz, A. (2023).
\newblock Components and their failure rates in autonomous driving.
\newblock In {\em Proceeding of the 33rd European Safety and Reliability
  Conference}, pages 233--240.

\bibitem[Saparia et~al., 2021]{saparia2021active}
Saparia, S., Schimpe, A., and Ferranti, L. (2021).
\newblock Active safety system for semi-autonomous teleoperated vehicles.
\newblock In {\em 2021 IEEE Intelligent Vehicles Symposium Workshops (IV
  Workshops)}, pages 141--147. IEEE.

\bibitem[Waymo, 2025]{Waymo2025DC}
Waymo (2025).
\newblock Next stop for waymo one: Washington, d.c.
\newblock Accessed: 2025-09-12.

\bibitem[Xiao et~al., 2019]{xiao_driving_2019}
Xiao, Y., Krunz, M., Volos, H., and Bando, T. (2019).
\newblock Driving in the {Fog}: {Latency} {Measurement}, {Modeling}, and
  {Optimization} of {LTE}-based {Fog} {Computing} for {Smart} {Vehicles}.
\newblock In {\em 2019 16th {Annual} {IEEE} {International} {Conference} on
  {Sensing}, {Communication}, and {Networking} ({SECON})}, pages 1--9, Boston,
  MA, USA. IEEE.

\bibitem[Zhao et~al., 2024]{zhao_remote_2024}
Zhao, L., Nybacka, M., Aramrattana, M., Rothhämel, M., Habibovic, A., Drugge,
  L., and Jiang, F. (2024).
\newblock Remote {Driving} of {Road} {Vehicles}: {A} {Survey} of {Driving}
  {Feedback}, {Latency}, {Support} {Control}, and {Real} {Applications}.
\newblock {\em IEEE Transactions on Intelligent Vehicles}, 9(10):6086--6107.

\bibitem[Zoox, 2025]{Zoox2025Vegas}
Zoox (2025).
\newblock Zoox opens its las vegas robotaxi service to the public.
\newblock Accessed: 2025-09-12.

\end{thebibliography}
}
\end{document}